\documentclass[twocolumn,times,tighten]{aastex631}

\usepackage{amsmath,amstext}
\usepackage[english]{babel}
\usepackage[utf8x]{inputenc}
\usepackage{textgreek}

\newcommand{\teff}{$T_\mathrm{eff}$}

\newcommand{\logg}{$\log g$}
\newcommand{\feh}{[Fe/H]}

\newcommand{\mic}{$\mu \mathrm m$}

\begin{document}

\title{Composition of Giants 1$^\circ$ North of the Galactic Center:\\
Detailed Abundance Trends for 21 Elements Observed with IGRINS}

\correspondingauthor{G. Nandakumar}
\email{govind.nandakumar@astro.lu.se}

\author[0000-0002-6077-2059]{Govind Nandakumar}
\affil{Division of Astrophysics, Department of Physics, Lund University, Box 43, SE-22100 Lund, Sweden}

\author[0000-0001-6294-3790]{Nils Ryde}
\affil{Division of Astrophysics, Department of Physics, Lund University, Box 43, SE-22100 Lund, Sweden}

\author[0000-0001-7875-6391]{Gregory Mace}
\affil{Department of Astronomy and McDonald Observatory, The University of Texas, Austin, TX 78712, USA}

\author[0000-0001-6909-3856]{Kyle F. Kaplan}
\affil{Department of Astronomy and McDonald Observatory, The University of Texas, Austin, TX 78712, USA}

\author[0000-0002-4080-2927]{Niels Nieuwmunster}
\affil{Division of Astrophysics, Department of Physics, Lund University, Box 43, SE-22100 Lund, Sweden}
\affil{Observatoire de la C\^ote d'Azur, CNRS UMR 7293, BP4229, Laboratoire Lagrange, F-06304 Nice Cedex 4, France}

\author[0000-0003-3577-3540]{Daniel Jaffe}
\affil{Department of Astronomy and McDonald Observatory, The University of Texas, Austin, TX 78712, USA}

\author[0000-0003-0427-8387]{R. Michael Rich}
\affil{Department of Physics and Astronomy, UCLA, 430 Portola Plaza, Box 951547, Los Angeles, CA 90095-1547, USA}

\author[0000-0002-6590-1657]{Mathias Schultheis}
\affil{Observatoire de la C\^ote d'Azur, CNRS UMR 7293, BP4229, Laboratoire Lagrange, F-06304 Nice Cedex 4, France}

\author[0000-0002-4287-1088]{Oscar Agertz}
\affil{Division of Astrophysics, Department of Physics, Lund University, Box 43, SE-22100 Lund, Sweden}

\author[0000-0003-3479-4606]{Eric Andersson}
\affil{Department of Astrophysics, American Museum of Natural History, 200 Central Park West, New York, NY 10024, USA}

\author[0000-0002-3456-5929]{Christopher Sneden}
\affil{Department of Astronomy and McDonald Observatory, The University of Texas, Austin, TX 78712, USA}

\author{Emily Strickland}
\affil{Department of Physics and Astronomy, The University of Utah, Salt Lake City, UT 84112, USA}

\author[0000-0002-5633-4400]{Brian Thorsbro}
\affil{Observatoire de la C\^ote d'Azur, CNRS UMR 7293, BP4229, Laboratoire Lagrange, F-06304 Nice Cedex 4, France}



\begin{abstract}





We report the first high resolution, detailed abundances of 21 elements for giants in the Galactic bulge/bar within $1^\circ$ of the Galactic plane, where high extinction has rendered such studies challenging. Our high S/N and high-resolution, near-infrared spectra of 7 M giants in the inner-bulge, located at ($l,b$)=(0,+1$^{\circ}$), are observed using the IGRINS spectrograph. We report the first multi-chemical study of the inner Galactic bulge, by investigating relative to a robust new Solar Neighborhood sample, the abundance trends of 21 elements, including the relatively difficult to study heavy elements. The elements studied are:  F, Mg, Si, S, Ca, Na, Al, K, Sc, Ti, V, Cr, Mn, Co, Ni, Cu, Zn, Y, Ce, Nd, and Yb. 

We investigate bulge membership of all seven stars using distances and orbital simulations, and find that the most metal-poor star may be a halo interloper. Our investigation shows that the inner-bulge also as close as $1^\circ$ North of the Galactic Center displays a similarity to the inner disk sequence, following the high [$\alpha$/Fe] envelope of the Solar vicinity metal-rich population, though no firm conclusions for a different enrichment history are evident from this sample. We find a small fraction of metal-poor stars (\feh$<-0.5$) but most of our stars are mainly of super-solar metallicity. Fluorine is found to be enhanced at high metallicity compared to the solar neighbourhood, but confirmation with a larger sample is required.  We will apply this approach to explore populations of the Nuclear Stellar Disk and the Nuclear Star Cluster. 
\end{abstract}


\section{Introduction}
\label{sec:intro}

Our canonical view of the chemical evolution of the Galactic bulge is grounded in a stellar population of age $>10$ Gyr that formed early and rapidly.  \citet{McWilliam:1994} found the bulge enhanced in alpha-elements relative to the thick disk; \citet{Matteucci:1990} illustrated the expected elevated trend of $[\alpha/\rm Fe]$ versus [Fe/H].  Subsequent modeling of chemical evolution of the bulge adds detail \citep[e.g.][]{Matteucci:2019} and posits a multi-phase evolution of the bulge, with the possibility of a disk-like population forming, following a pause in star formation after the early starburst that gives rise to the bulge. The picture of alpha-enhancement relative to the thin disk has largely held firm \citep[e.g.][]{Zoccali:2006,Fulbright:2007,melendez:08,ryde:10,
Hill:2011,johnson:14,jonsson:17,bensby:17,lomaeva:19,Nieuwmunster:2023} and a picture is emerging of the bulge appearing to be similar to the thick disk, but extended to higher metallicity.  This was also suggested by \cite{Dimatteo:2016}, where the thick-disk (high $\alpha$) and the metal-poor thin-disk (low $\alpha$, \feh$\lesssim$-0.3) stellar populations in the the solar neighbourhood are referred to as the ‘inner disk sequence’ and the ‘outer disk sequence’ respectively. Further, the metal-rich thin-disk population that is joined by the ‘inner disk sequence’ is referred to as the ‘inner thin disk’ and it is considered to be the same structure as the thick-disk or ‘inner disk sequence’ \citep[see also][]{Haywood:2013}. Based on this, the metal-rich component (A) and metal-poor components (B, C) in the bulge from \cite{Ness:2013} are associated with the ‘inner thin disk’ and ‘inner disk sequence’ sequences respectively. A similar scenario is borne out in the cosmological zoom-simulation  {\small VINTERGATAN} \citep{Agertz2021,Renaud2021} of a Milky Way-like galaxy, where the solar neighbourhood chemical thin disk stars formed in an outer disk.


Industrial-scale spectroscopy of the bulge and disk by the APOGEE project \citep[e.g.][]{hayden:15,SH:20,Queiroz:2021} finds a bifurcated trend in $[\alpha/\rm Fe]$ versus [Fe/H] as a function of radius in the Galaxy, with the inner 2 kpc dominated by a high alpha population and disk-like, low-alpha population more or less disconnected, in contrast to the continuous trends in  $[\alpha/\rm Fe]$ versus [Fe/H] found earlier. Meanwhile, \cite{Zasowski:2019} found the lack of a bimodality in the [$\alpha$/Fe] distributions at subsolar metallicity suggesting that the abundances are dominated by a single chemical evolutionary sequence using $\sim$ 4000 K/M giants in APOGEE DR14/DR15 that are at galactocentric distances, R $\leq$ 4.0 kpc.

We stand on the threshold of vastly larger production of industrial scale spectroscopy by the MOONS project and SDSS-V.  The former will, like APOGEE, employ the infrared H-band to report compositions for $\sim$ 15 elements at R$\sim$ 20,000 and do so, ultimately, for millions of stars.  

As studies of the bulge/bar/inner galaxy increase, it has become appreciated that its star formation history may be more complex \citep{bensby:17} and even re-assessments of age \citep{Joyce:2023} support an age range among the most metal rich stars.  \cite{Johnson:2022} finds a striking concentration of bulge/bar metal rich stars to the Galactic plane.  Unfortunately, the precision and breadth of spectroscopy has failed to keep pace with these advances.  

The inner bulge regions, i.e, regions within $|b|< 2^{\circ}$, host the Milky Way nuclear stellar disk (NSD) and nuclear star cluster (NSC), that are also ubiquitous in other galaxies \citep{Neumayer:2020}. But detailed investigations into the chemistry of different stellar populations in the inner bulge regions have been limited to a handful of spectroscopic observations with only limited number of stars analysed in the region (for details, see the introduction and references in \cite{Nandakumar:18}). This is due to the crowding and dust obscuration along the line-of-sight toward the inner bulge that demands observations at infrared wavelengths. Thus, more near-infrared spectroscopic observations are necessary to decipher the chemistry of the inner bulge regions.

Here we report on a small sample of bulge/bar candidate giants near the Galactic plane that are compared with an identically observed sample of Solar neighborhood stars, using the Immersion GRating INfrared Spectrograph \citep[IGRINS;][]{Yuk:2010,Wang:2010,Gully:2012,Moon:2012,Park:2014,Jeong:2014}. The bulge/bar candidate giants are chosen from a field along the northern minor axis at ($l = 0^\circ$, $b = +1^\circ$), which is a region with relatively low extinction that is observable with the 2.7 meter (107-inch) Harlan J. Smith Telescope at McDonald Observatory. IGRINS is able to record both the infrared H and K bands (1.45 - 2.5 $\mu$m) simultaneously at $R=45,000$, double the resolution of the "APOGEE-like" surveys, and allows for the detailed analysis of a wealth of spectral lines enabling a detailed study of stellar abundances for a range of different elements. Although the sample we report here is small, we have the advantage of a direct comparison with the Solar neighborhood sample, and we can investigate the detailed abundances of 21 elements for these stars only $1^\circ$ from the Galactic plane.  As has already been hinted at by \cite{Nieuwmunster:2023}, differences between this population and the thick disk are subtle at best. 
The aim of this study -- the careful, detailed analysis of our new sample compared to the \cite{Nandakumar:2023} solar vicinity study -- is to provide a well established baseline for the study of chemical evolution of this region, and to make clear the observational challenges as this work proceeds to larger samples.  Finally, this work will stand in contrast to the future industrial scale samples that will emerge in the coming years.

In this work, we will thus determine elemental abundances and their trends with metallicities for up to 21 elements for 7 M-giants located in inner degree region of the Galactic Center along the Northern minor axis ($l = 0^{\circ}$, $b = +1^{\circ}$). These abundances are derived based on a consistent and systematic spectroscopic analysis of IGRINS spectra. The details of the observations and data reduction are provided in the section~\ref{sec:obs} followed by the details of the spectroscopic and dynamic analysis in the section~\ref{sec:analysis}. In section~\ref{sec:results}, we show the elemental abundance trends for different groups of elements ($\alpha$, odd-Z, iron-peak, and neutron-capture elements) and also compare with the trends for the same group of elements for the 50 solar-neighborhood stars, observed and analysed with the same setup. 
We discuss our results in the context of the Galactic bulge and inner Milky Way structures, and compare with inner Milky Way abundance trends from APOGEE in section~\ref{sec:discussion}. We make our final conclusions in section~\ref{sec:conclusion}.

\begin{table*}
\caption{ Observational details of the seven M-giants.}\label{table:obs}
\centering
\begin{tabular}{c c c c c c c c c c}
\hline
\hline
Star & RA & Dec  & Date  & J$_\mathrm{2MASS}$  & H$_\mathrm{2MASS}$    & K$_\mathrm{2MASS}$    & SNR$_\mathrm{H}$ & SNR$_\mathrm{K}$ \\

    &   & & UT &  (mag) &  (mag) & (mag) &    \multicolumn{2}{c}{(per resolution element)} \\
\hline
BP1$\_$1 & 17:41:49.98 & -28:27:32.70   & 2016-05-23               & 12.02 &  10.57 & 9.99   &     50   &     70     \\  
BP1$\_$2 & 17:41:56.10 & -28:23:36.68   & 2016-05-24               & 12.28 & 10.73   & 10.02  &   50  & 50              \\  
BP1$\_$3 & 17:41:44.96 &-28:28:06.37     & 2016-05-28              & 11.42 & 10.12  & 9.67   &  80  &    90           \\  
BP1$\_$5 & 17:41:56.59 &-28:25:37.60     & 2016-06-18              & 11.69 & 10.24 & 9.70   &  60  & 60              \\  
BP1$\_$6 & 17:41:27.03 &  -28:22:49.23 & 2016-06-19 and 2016-07-05 & 11.62 & 10.13 & 9.54   &  130  &  110          \\  
BP1$\_$7 & 17:41:38.81 & -28:28:39.89   & 2016-06-20               & 11.74 & 10.49  &  9.52   &  90  &  70         \\  
BP1$\_$9 & 17:41:36.94 & -28:23:41.07    & 2016-07-06              & 11.96 & 10.38  &  9.72   &   90  &     80       \\    

\hline
\hline                                 
\end{tabular}
\end{table*}

\section{Observations and Data reduction}
\label{sec:obs}

\begin{figure}
  \includegraphics[width=\columnwidth]{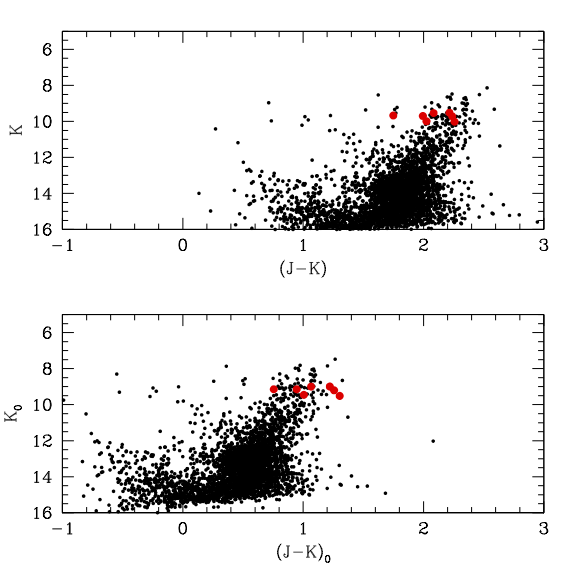}
  \caption{ K versus J-K color magnitude diagram without extinction correction (top panel) and with extinction correction (bottom panel) for the stars in the $l = 0^{\circ}$, $b = +1^{\circ}$ field. The seven inner-bulge giants observed and analysed in this work are shown in red circles. }
  \label{fig:CMD}%
\end{figure}

In order to investigate the chemical compositions and elemental trends of 
stars close to the Galactic plane,
we have observed the entire H- and K-bands (1.5 - 1.75 and 2.05 - 2.3 \mic, respectively) of 9 M giants with IGRINS at high spectral-resolution, $R\sim45,000$. 
The observations were carried out with the $2.7$~meter (107-inch) Harlan J. Smith Telescope at McDonald Observatory in May-July 2016, see Table \ref{table:obs}. However, only those 7 stars that have high enough signal-to-noise ratios ($\gtrsim50$  per pixel) for a precise abundance analysis, will be used in the following abundance analysis. We show the location of these seven stars (red circles) in the K versus J-K color magnitude diagram with and without extinction correction in the  bottom and top panels of Figure~\ref{fig:CMD}, respectively. As a reference sample we use the 50 M-giants in the solar neighbourhood analysed in \cite{Nandakumar:2024}, \cite{Nandakumar:2023} and \cite{Nandakumar:2023b}. These stars were observed with IGRINS in the same way as the inner-bulge stars presented here.



The spectral reductions were done with the IGRINS PipeLine Package \citep[IGRINS PLP;][]{Lee:2017} to optimally extract the telluric corrected, wavelength calibrated spectra after flat-field correction \citep{Han:2012,Oh:2014}. The spectra were then resampled and normalized in {\tt iraf} \citep{IRAF} but to take care of any residual modulations in the continuum levels, we put considerable focus on defining specific local continua around every line that elemental abundances were determined from. Finally, the spectra are shifted to laboratory wavelengths in air after a stellar radial velocity correction. The average signal-to-noise ratios (SNR)\footnote{SNR is provided by RRISA \citep[The Raw $\&$ Reduced IGRINS Spectral Archive;][]{rrisa} and is the average SNR for K band and is per resolution element. It varies over the orders and it is lowest at the ends of the orders} of the spectra of the giants close to the plane are in the range $60-130$ (see Table \ref{table:obs}).

\begin{table*}
\caption{Stellar parameters, Galactocentric radial velocities and bulge membership of the seven M-giants.}\label{table:parameters}
\begin{tabular}{c c c c c c c c c c c c}
\hline
\hline
 Star   & $T_\mathrm{eff}$ & $\log g$  & [Fe/H]  &  $\xi_\mathrm{micro}$ & [C/Fe] & [N/Fe] & [O/Fe] & V$_{R}$ & population \\
  & K & log(cm/s$^{2}$) & dex & km/s & dex & dex & dex & kms$^{-1}$ & \\
  \hline
BP1$\_$1    &  3491  &  0.79  &  0.21  &  1.72  &  0.11  &  0.05  &  0.04  & 41.14 & inner-bulge  \\  
BP1$\_$2    &  3455  &  0.6  &  0.03  &  2.13  &  0.16  &  0.19  &  0.14   & -64.69 & inner-bulge \\  
BP1$\_$3    &  3995  &  1.68  &  0.18  &  1.59  &  0.06  &  0.22  &  0.06  & -44.42 & inner-bulge  \\  
BP1$\_$5    &  3956  &  0.69  &  -1.14  &  1.67  &  -0.12  &  0.53  &  0.6 & -94.94 & inner halo   \\  
BP1$\_$6    &  3739  &  1.24  &  0.23  &  1.85  &  0.09  &  0.29  &  0.04  & -78.95 & inner-bulge  \\  
BP1$\_$7    &  3743  &  0.7  &  -0.65  &  1.8  &  0.16  &  0.18  &  0.44   & 5.84 & inner-bulge  \\  
BP1$\_$9    &  3460  &  0.62  &  0.06  &  2.23  &  0.12  &  0.16  &  0.13  & 205.70 & inner-bulge  \\    

\hline

\hline
\hline
\end{tabular}
 
\end{table*}

\section{Analysis}
\label{sec:analysis}

In order to minimize any systematic uncertainties, 
measurements of the stellar abundances need to be carried out differentially with respect to the stellar populations in 
the solar neighborhood. 
It is then imperative to observe stars of similar spectral types (here, M giants) both in the inner Milky Way and in the solar neighborhood regions, using the same instrument with the same wavelength range and spectral resolution.  
Furthermore, the same pipelines and techniques must be used in the determination of the stellar parameters and elemental abundances. 


\subsection{Spectroscopic analysis}
\label{sec:parameters}

\cite{Nandakumar:2024} determined abundances of the same 21 elements from the exact same lines for 50 
close-by M giants ($< 4000$~K) also from high-resolution spectra observed with IGRINS in the same mode. They demonstrated that M giants observed at high spectral resolution in the H and K bands ($1.5-2.4$\,\mic) can yield both accurate and precise 
abundance-ratio trends versus metallicity for all these elements. The 50  solar-neighbourhood stars will serve as the reference sample here.

Similar to \cite{Nandakumar:2024}), the spectroscopic analysis of the seven inner-bulge M giants in this work has been carried out with the spectral synthesis method using the spectroscopy made easy \citep[SME;][]{sme,sme_code} tool in combination with the grid of one-dimensional (1D) Model Atmospheres in a Radiative and Convective Scheme (MARCS) stellar atmosphere models \citep{marcs:08}. We use the method provided by \citet{Nandakumar:2023} to determine  the stellar parameters for the inner-bulge stars. In this method the effective temperature is mainly constrained by selected \teff-sensitive OH lines, the metallicity by selected Fe lines, the microturbulence by different sets of weak and strong lines, and the C and N abundances by selected CO and CN molecular lines. The surface gravity is determined based on the effective temperature and metallicity by means of the Yonsei-Yale (YY) isochrones assuming old ages of 2-10 Gyr \citep{Demarque:2004}, which is appropriate for low-mass giants. This is the same method used to determine stellar parameters for the reference sample of 50 solar neighborhood M giants. The method and its validation with nearby benchmark stars are described in detail in \cite{Nandakumar:2023}. We note that there also exists an excellent method to determine the effective temperature through a Line-Depth Method for IGRINS spectra as presented in \citet{afsar:2023}. We chose, however, to get {\it all} the stellar parameters through spectral synthesis.

The oxygen abundances for the inner Bulge stars have been assumed to follow the enhanced values of the thick-disk population based on findings from previous studies \citep[for example, see][]{Rojas-Arriagada:2017,Nieuwmunster:2023}. Hence, during parameter determination of the seven inner bulge stars, we assumed oxygen abundances (corresponding to the metallicity of each star) from the thick-disk [O/Fe] versus [Fe/H] trend determined for solar neighborhood stars in \cite{Amarsi:2019} (see also the simple functional form of this trend in Figure 1. in \citealt{Nandakumar:2023}). The effective temperatures of the seven inner Bulge stars lie in the range of 3500 to 4000 K, and the surface gravities in the range of 0.6 to 1.7 dex. Two stars are metal poor (BP1$\_$5: \feh$\sim$ -1.14 dex and BP1$\_$7: \feh$\sim$ -0.65 dex), and the rest are all metal rich or super-solar (\feh$\sim$ 0.03-0.23 dex). Our final stellar parameters, [C/Fe], [N/Fe], and [O/Fe] results, determined in this way, are listed in the Table~\ref{table:parameters}.

\begin{figure*}
  \includegraphics[width=\textwidth]{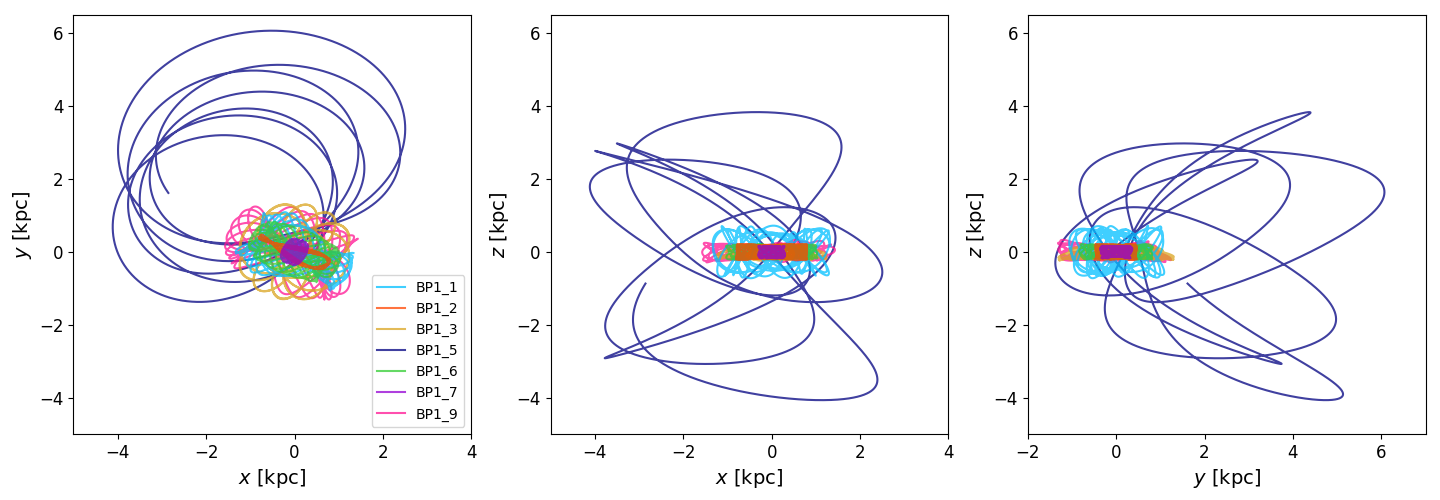}
  \caption{Orbits of the seven M-giants of the ($l,b$)=(0,+1$^{\circ}$) field in the $(x,y)$ \textit{(left panel)}, $(x,z)$ \textit{(middle panel)} and $(y,z)$ \textit{(right panel)}.}
  \label{fig:orbits}%
\end{figure*} 

\begin{figure}
  \includegraphics[width=\columnwidth]{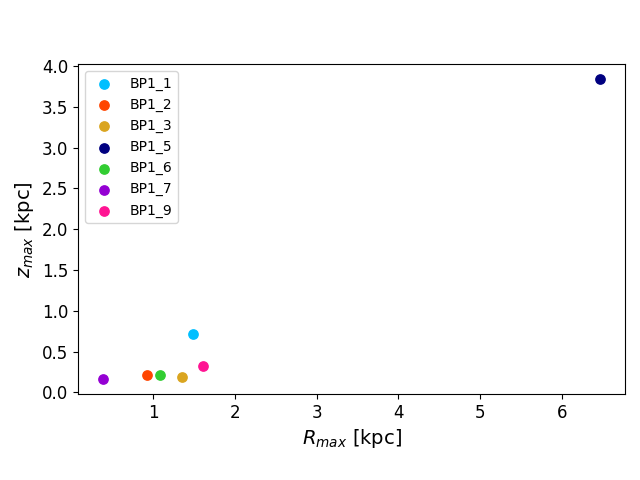}
  \caption{Maximal height from the Galactic plane $z_{max}$ vs. the apocentric radius $R_{max}$ diagram for the seven M-giants of the the ($l,b$)=(0,+1$^{\circ}$) field.}
  \label{fig:rmax_zmax}%
\end{figure}

We will use the method outlined in \citet{Nandakumar:2023} and \cite{Nandakumar:2024} to determine the elemental abundances for the 21 elements and Fe. The atomic lines used are given in these references and the abundances are calculated assuming non-local thermodynamic equilibrium (NLTE) for C, N, O, Na, Mg, Al, Si, K, Ca, and Fe (\citealt{NLTE}, \citealt{lind17} and \citealt{amarsi16} with subsequent updates; Amarsi priv. comm.), with the departure coefficients computed using the MPI-parallelized NLTE radiative transfer code \texttt{Balder} \citep{Amarsi:2018}. The line data for the CO, CN, and OH lines were adopted from the line lists of \citet{li:2015}, \citet{brooke:2016}, and \citet{sneden:2014}, respectively. The molecular lines are important since they blend with and affect the measurement of atomic spectral lines.

\subsection{Stellar parameters with thin-disk Oxygen assumption}
\label{sec:thin_parameters}

As mentioned in the previous section, the main assumption in the parameter determination method is the oxygen abundance, [O/Fe], that is assigned an abundance value based on a thin- or thick-disk metallicity dependent trend, following the [O/Fe] versus [Fe/H] trends in \cite{Amarsi:2019}. 
We have assumed that the inner-bulge stars follow the thick-disk [O/Fe] trend based on several previous studies, while determining stellar parameters. Here we carry out the exercise where we assume thin-disk [O/Fe] abundances instead, and redetermine the stellar parameters, even though there are no reason to believe that this should be the case. This leads to decrease in [O/Fe] by 0.05 to 0.35 dex which in turn results in the decrease in \teff\, by 50 to 150 K, \logg\, by 0.1 to 0.25 dex, \feh\, by 0.0 to 0.15 dex, and $\xi_\mathrm{micro}$ by -0.05 to 0.2 km/s. The most metal-poor star, BP1$\_$7 and BP1$\_$5, have largest differences in [O/Fe] of 0.19 and 0.35 dex respectively, and thus largest differences in \teff\, (100 K and 150 K) and \logg\, ($\sim$0.25 dex) as well.

\cite{Nandakumar:2023} showed that when the thin- or thick- disk population assumption is wrong, the Mg abundances determined based on the resulting stellar parameters can be used to catch any such misclassification. This is illustrated in the Figure 7 in \cite{Nandakumar:2023} where the [Mg/Fe] abundances decrease further for the incorrectly classified thin-disk star and it increases further for the incorrectly classified thick-disk star. The Mg abundances determined for the six inner-bulge stars and the halo interloper from the new sets of stellar parameters, with the assumption of [O/Fe] following thin-disk, are enhanced by 0.02-0.05 dex, and thus further confirms our assumption of a thick-disk trend for the [O/Fe] vs \feh\, while determining the stellar parameters. 


\subsection{Orbital analysis}
\label{sec:dynamic}

To confirm that the stars observed in the ($l,b$)=(0,+1$^{\circ}$) field belong to the bulge, we performed orbit calculations. We used the software package AGAMA \citep{agama} to integrate the orbit over 500 Myr. The gravitational potential used is composed of three components: a halo, a disc, both from \texttt{MWPotential14} \citep{galpy} and a bar/bulge from \cite{launhardtbar}. The latter, which is initially at an angle of $\alpha = 25^{\circ}$ from the line of sight towards the Galactic centre, is rotating clockwise with a pattern speed about $\Omega_{b}= 40$\,km s$^{-1}$ kpc$^{-1}$ \citep{Portail17}. Among the initial conditions (i.e. position, proper motions, radial velocity, and distance) necessary to integrate the orbits, the position and proper motions come from the Gaia EDR3 survey (\citealp{gaia2016,gaia2021}) and radial velocities determined based on the wavelength shift of known lines in the observed spectra with respect to their wavelengths in vacuum. We estimated the spectro-photometric distances as explained by \cite{distance1} and \cite{Schultheis2017} using the stellar parameters \teff, \logg, and \feh\ and the near-infrared photometry J, H, and K and associated errors. The coordinates of the orbits as a function of time and the orbital parameters like the apocentric radius $R_{max}$ and the maximum height from the Galactic plane $z_{max}$, allow us to determine the bulge membership.

\section{Results}
\label{sec:results}

In Figure \ref{fig:orbits} we plot the orbits for the seven stars observed at ($l,b$)=(0,+1$^{\circ}$). All stars except one are clearly members of the inner-bulge and stay confined within a kpc of the Galactic center. The orbit calculation of the star BP1\_5, however, shows that it more likely is a halo star passing by the center.  We plot the maximum height from the Galactic plane $z_{max}$ versus the apocentric radius $R_{max}$ in Figure \ref{fig:rmax_zmax}. In this Figure it is also obvious that we should exclude this star (BP1\_5) from the discussion of the chemistry of stars in the inner-bulge. The metallicity of this star (\feh=-1.14), is also more compatible of it being a halo star than an inner-bulge stars \cite[see, for example, the discussion on the metallicity distribution of the inner-bulge in][]{Rojas-Arriagada:2020}.  We have therefore marked this metal-poor star separately in the following discussion on the abundances versus metallicity.


We have thus determined the metallicities and abundances of 21 elements for the six inner-bulge stars, one halo interloper, and the sample of 50 M giants in the solar neighborhood. For all elements except fluorine and cerium, we have scaled the abundances with respect to solar abundance values from \cite{solar:sme}. We used A$_{\odot}$(F) = 4.43 from \cite{Lodders:2003} and A$_{\odot}$(Ce) =  1.58 from \cite{Grevesse:2015} for fluorine and cerium respectively. 

We classified the solar neighborhood M-giants as belonging to the chemically defined thin- and thick-disk populations based on their enhanced APOGEE magnesium abundance from the [Mg/Fe] versus [Fe/H] plot and confirmed by our magnesium abundance determinations. In the following sections, we adopt the ideology from \cite{Haywood:2013} while labeling the solar neighborhood stars as thin- and thick-disk, i.e., when we mention thick-disk and metal-rich thin disk we refer to the inner disk sequence.

\begin{figure*}
  \includegraphics[width=\textwidth]{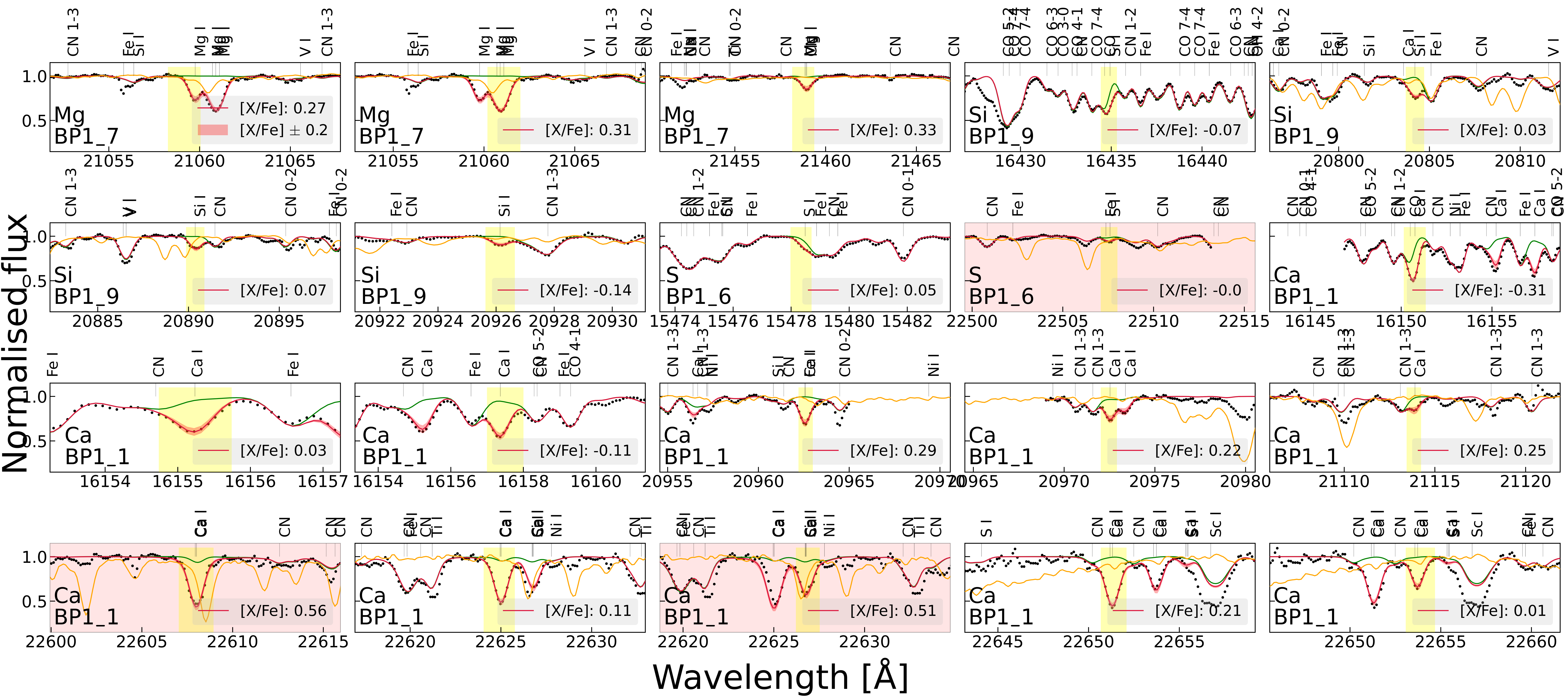}
  \caption{ Wavelength regions centered at the three selected magnesium lines for the star BP1$\_$7, four selected silicon lines for the star BP1$\_$9, two selected sulfur lines for the star BP1$\_$6, and 11 selected calcium lines for the  star BP1$\_$1. In each panel, the black circles denote the observed spectrum, the crimson line denotes the best-fit synthetic spectrum, and the red band denotes the variation in the synthetic spectrum for a difference of $\pm$0.2\,dex in the abundances. The yellow bands in each panel represent the line masks defined for each line, wherein SME fits observed spectra by varying the elemental abundance and finds the best synthetic spectra fit by $\chi^{2}$ minimization. The green line shows the synthetic spectrum without the absorption from the element, also indicating any possible blends in the line. The abundance values corresponding to the best-fit case for each line are listed in each panel. All identified atomic and molecular lines are also denoted in the top of each panel. The panels with red background correspond to the lines we have avoided due to bad noisy spectra or strong telluric contamination.  }
  \label{fig:alphaspectra}%
\end{figure*}

\begin{figure*}
  \includegraphics[width=\textwidth]{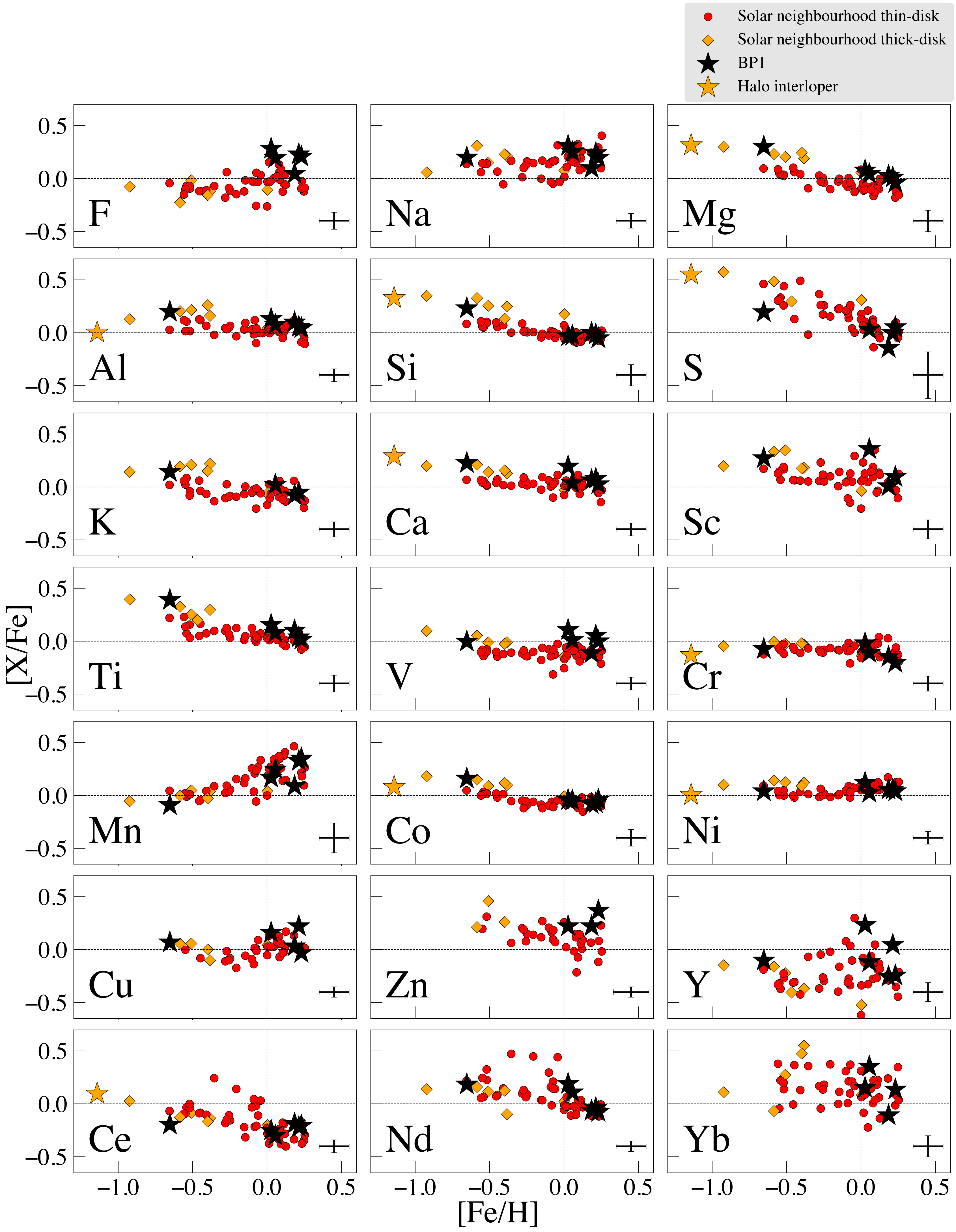}
  \caption{ Elemental abundances of 21 elements versus metallicity for 50 solar neighbourhood thin- (red filled circles) and thick-disk stars (orange diamonds), six inner-bulge stars (black stars), and the halo interloper (orange star). Elemental abundances are mean values determined from multiple lines for all elements except V, Co, Zn, and Yb.}
  \label{fig:all_gild_trend}%
\end{figure*}

We show the observed and synthetic spectra of selected inner-bulge stars for the absorption lines used in this work to determine abundances of magnesium, silicon, sulfur, and calcium in the Figure \ref{fig:alphaspectra}. The abundance trends are shown in Figure \ref{fig:all_gild_trend}, with the inner-bulge stars marked with black star symbols, the halo interloper with an orange star symbol, and the reference sample in red circles (low-alpha or thin-disk stars) and orange diamonds (high-alpha  or thick-disk stars). 

We estimated the uncertainties in the elemental abundances, shown as error bars in Figure~\ref{fig:all_gild_trend}, for the star BP1$\_$9. We redetermined 50 values of the abundances from all selected lines of each element by setting the stellar parameters to values randomly chosen from normal distributions with the actual stellar parameter value (listed in the Table~\ref{table:parameters}) as the mean and the typical uncertainties ($\pm$100 K in \teff, $\pm$0.2 dex in \logg, $\pm$0.1 dex in \feh, and $\pm$0.1 km s$^{-1}$ in $\xi_\mathrm{micro}$) as the standard deviation. The uncertainty in the abundance determined from each line is then taken to be the mid-value of the difference between the 84$^{th}$ and 16$^{th}$ percentiles of the abundance distribution. The final abundance uncertainty is the standard error of mean of uncertainties from all chosen lines of an element. The estimated uncertainties are in the range of 0.05 - 0.14 dex for all elements except sulfur for which we estimate a larger uncertainty of 0.2 dex. We carried out the same exercise for a typical solar metallicity star in the solar neighbourhood reference sample and we get similar abundance uncertainties.     

The metallicities of the stars were determined as one of the stellar parameters. Although we only have very few stars in the inner-bulge set, their distribution in metallicity fits well in the emerging picture from \citet{Schultheis:2015,ryde:2016_bp2,Nandakumar:18,Nieuwmunster:2023}. Five of our stars fit the metal-rich peak, and the metal-poor star fits within the sub-solar peak found in these other studies too. The metal-rich peak increases in relative terms, getting progressively dominant,  the closer to the plane the field lies \citep{Nandakumar:18}.

Next, we will compare the element ratio trends versus metallicity for the 21 elements determined for the inner-bulge population and compare them differentially with those of the reference sample of stars in the local disks.

\subsection{Fluorine (F)}
\label{sec:fluorine}

{\bf Fluorine.} The fluorine abundances are determined from 6 HF molecular lines in the K band, see \cite{Nandakumar:2023b}.
For the solar neighbourhood sample \citep[also presented in][]{Nandakumar:2023b}, we find a flat [F/Fe] versus [Fe/H] trend, but a hint of a slightly higher level of [F/Fe] for stars with \feh\,$>$ 0 dex. In comparison, the five metal-rich, inner-bulge stars have even higher derived [F/Fe] abundances. For stars with several useful lines, all the lines yield elevated F abundances, ensuring a precise derived mean abundance. 
The spectra are of good quality for all lines and they are all abundance-sensitive and therefore not saturated. Our finding is therefore interesting, but due to the temperature sensitivity of the lines, more stars are needed in order to draw any firm conclusions. No reliable fluorine abundances could be derived for the metal-poor, inner-bulge star due to noisy spectra and the fact that telluric lines severely affected the HF lines for this star.









\subsection{$\alpha$ elements (Mg, Si, S, and Ca)}
\label{sec:alpha}

{\bf Magnesium, silicon, and calcium} 
There is a clear enhancement in the mean abundances of the three $\alpha$ elements, Mg, Si, and Ca for the solar-neighbourhood, thick-disk stars leading to a separation between the thin- and thick-disk populations. 
The metal-poor, inner-bulge star shows similar enhancements to those shown by the thick-disk stars for these $\alpha$ elements. The rest of the 5 metal-rich, inner-bulge stars also have enhanced abundances in the case of Mg and Ca  compared to the metal-rich, thin-disk stars and lie close to the upper envelope of the thin-disk abundance trend. In the case of Si, the mean abundances for the metal-rich, inner-bulge stars overlap with the metal-rich, thin-disk abundance trend. Overall, we find clear enhancements in the $\alpha$-element abundances for the inner-bulge stars that are consistent with the enhancement seen for the thick-disk stellar population.

{\bf Sulfur.} The sulfur abundance determination is hampered with large uncertainties and therefore a large scatter in the abundance trends. We could only determine reliable abundance from one weak H-band line for the inner-bulge stars since the K-band line is severely affected by telluric lines. The metal-poor, halo interloper star (BP1$\_$5) shows an expected  elevated  [S/Fe] abundance ratio, whereas the  metal-poor, inner-bulge star shows a lower sulfur abundance. As mentioned earlier the uncertainties in the sulfur determination ($\pm 0.2$\,dex) is, however, largest of all the elements. The scatter in abundances for the solar neighbourhood stars is also more prominent at lower metallicities (see also Figure~\ref{fig:all_apogee_trend}). We would need a larger sample of inner-bulge stars in this metallicity regime to discern whether it follows the thick disk or not.


\subsection{Odd-Z elements (Na, Al, and K)}
\label{sec:oddz}


{\bf Sodium.}  
The mean [Na/Fe] trend for the solar neighbourhood sample resembles 
a stretched out N-shape \citep[c.f.][]{bensby:2014}. The trend for the metal-poor, thick-disk stars shows a hint of a separate trend compared to that of the thin-disk stars, following the upper envelope of the thin-disk trend. 
The [Na/Fe] values determined for the metal-poor, inner-bulge star (BP1$\_$7) cannot be differentiated from either disk-trends
in the solar neighbourhood. 
The five metal-rich, inner-bulge stars overlap with the solar-neighbourhood trend.  


{\bf Aluminum.}  
The mean [Al/Fe] of the solar-neighbourhood, thin-disk stars are generally near solar (in the super-solar regime).
The [Al/Fe] values of the thick-disk stars are enhanced by $\sim$ 0.2 dex, and decreases as metallicity decreases, which is also expected from chemical-evolution models \citep[see, for example,][]{kobayashi:20}. 
The mean [Al/Fe] for the metal-poor, inner-bulge star is consistent with the thick-disk trend. The mean [Al/Fe] values of the metal-rich, inner-bulge stars roughly follow the solar-neighbourhood, disk-population trend, perhaps with a tendency of following an upper envelope.

{\bf Potassium.}  
For the metal-poor, thick-disk stars there is a clear enhancement in [K/Fe]. The mean [K/Fe] values for the thin-disk stars are slightly sub-solar and show a near-straight trend. Only four inner-bulge stars have reliable [K/Fe] determined from good-quality lines in their spectra. Three of them are metal-rich and have mean abundance values at the upper envelope of the trend of solar-neighborhood, metal-rich, thin-disk stars. The enhanced abundance of the metal-poor, inner-bulge star (BP1$\_$7) is consistent with the enhanced values of the solar-neighborhood, metal-poor, thick-disk stars.

\subsection{Iron-peak elements (Sc, Ti, V, Cr, Mn, Co, and Ni)}
\label{sec:ironpeak}


{\bf Scandium.}  
The mean [Sc/Fe] trend in the solar-neighbourhood shows a generally decreasing slope with metallicity, with the thick-disk stars showing higher Sc abundances than those of the thin-disk.  
The metal-poor, inner-bulge star (BP1$\_$7) follow the thick-disk trend and the metal-rich ones follow the solar-neighbourhood, metal-rich, thin-disk trend. These show a similar spread in abundance for a given metallicity. These lines are temperature sensitive \citep[see, e.g.,][]{thorsbro:2018} which could be the cause of the large spread.

{\bf Titanium.}  
The thick-disk trend of [Ti/Fe] is clearly separated from that of the thin-disk for the solar-neighbourhood stars. 
The metal-poor, inner-bulge star (BP1$\_$7) shows a similar enhancement as the thick-disk stars. The rest of the 5 metal-rich, inner-bulge stars also have slightly enhanced abundances compared to the metal-rich, thin-disk stars and lie close to the upper envelope of the metal-rich, thin-disk abundance trend.

{\bf Vanadium.}  The thin-disk stars have sub-solar values of [V/Fe] and the metal-poor, thick-disk stars have enhanced super-solar values in the solar-neighbourhood, and shows a flat trend. The metal-poor, inner-bulge star (BP1$\_$7) shows a similar enhancement as the thick-disk stars. The rest of the 5 metal-rich, inner-bulge stars also have slightly enhanced abundances compared to the metal-rich, thin-disk stars and lie close to the upper envelope of the metal-rich, thin-disk abundance trend.

{\bf Chromium.} 
The mean [Cr/Fe] values for the solar-neighbourhood stars are mainly subsolar at all metallicities with larger scatter for metal-rich stars. Overall, this shows an expected flat trend for Cr \citep[e.g.][]{bensby:17}. The abundance differences between the thin- and thick-disk stars is small, although the metal-poor, thick-disk stars show a consistent enhancement in [Cr/Fe] compared to the thin-disk stars from all three lines used to determine abundances. 
The measured mean abundance for all six inner-bulge stars is consistent with the solar neighborhood trend. We cannot rule out that the metal-poor, inner-bulge star follow the thick-disk trend. 

{\bf Manganese.} Based on the  discussion in \cite{Nandakumar:2024}, showing that Mn empirically is more like a normal iron-peak element, there are reasons to believe that the [Mn/Fe] versus [Fe/H]-trend in the solar neighbourhood is essentially flat. The non-LTE corrections for the  M giants analysed here are, therefore, most likely  underestimated for the highest metallicities. However, when comparing the solar neighbourhood populations and the inner-bulge stars, we are only interested in the differential differences between the populations' trends and not so much its absolute values. The [Mn/Fe] trend  of the inner-bulge stars  nicely follows the solar neighbourhood one.

{\bf Cobalt.}  
The trend for the thin-disk, solar-neighbourhood stars for [Co/Fe] starts with super-solar values at lowest metallicities that transitions to subsolar values at [Fe/H]$\sim$ -0.4\,dex and continues at sub-solar values as the metallicity increases. All five metal-poor, thick-disk stars have enhanced [Co/Fe] values, and the solar metallicity thick-disk star has a solar [Co/Fe] value that is closer to the thin-disk stars.  The thin- and thick-disk trends are clearly separated. 
Similar enhancement in [Co/Fe] as for the thick-disk stars is found for the metal-poor, inner-bulge star (BP1$\_$7) while the five metal-rich, inner-bulge stars follow the metal-rich, thin-disk abundances.

{\bf Nickel.}  
The mean [Ni/Fe] values for the thin-disk stars show a tight trend that increases slightly with increasing metallicity. 
The thick-disk stars show enhancements with super-solar values. The [Ni/Fe] values for the metal-poor, inner-bulge star (BP1$\_$7) follow the disk trends and those of the five metal-rich, inner-bulge stars are consistent with the solar-neighborhood trend.

\subsection{Transition elements between iron-peak and neutron-capture elements (Zn)}


{\bf Zinc.} 
The abundance trend of [Zn/Fe] resembles an $\alpha$ element, decreasing with metallicity 
and with the thick-disk trend higher than that of the thin-disk. 
We could determine the [Zn/Fe] abundances for only three metal-rich stars in the inner-bulge, and they are consistent with the solar-neighborhood values in the super-solar metallicity range, tending to follow the upper envelope.

\subsection{Neutron-capture elements (Cu, Zn, Y, Ce, Nd, and Yb)}
We have determined the abundances of copper (Cu), yttrium (Y), cerium (Ce), neodymium (Nd), and ytterbium (Yb), which are all mainly synthesised by neutron-capture processes. Cu is synthesised through the weak-s process in massive stars \citep[e.g][]{Cu_ges:21,pagin:10,heitor:20}. Most neutron-capture elements are produced by a combination of the s- and r-processes.  For Y and Ce, the s-process dominates with s/r=70/30 and 85/15, respectively,  in the solar isotopic mixture \citep{bisterzo:14,prantzos:20}.  Neodymium has a 60/40 ratio and can therefore still be considered as a predominantly an s-process element. The more the r-process contributes the more the abundance trends will resemble and reveal an r-process contribution. On the other hand, 
ytterbium has a ratio close to 50/50 \citep{bisterzo:14,prantzos:20,kobayashi:20}, which makes Yb the most r-process rich element among the elements we are investigating. 


\subsubsection{Weak s-elements (Cu)}
\label{sec:weak}

{\bf Copper.} 
The mean [Cu/Fe]-abundances for the solar-neighbourhood stars show 
a wave-like trend, starting off at slightly supersolar values at the lowest metallicities, to then decrease to sub-solar values at sub-solar metallicities, but then again increase at super-solar metallicities. The thick-disk stars tend to lie above that of the thin-disk. 

The inner-bulge stars follow these trends nicely, following the thick-disk and then being consistent with the  solar-neighborhood [Cu/Fe] values at supersolar metallicities. 



\subsubsection{s-process dominated elements (Y, Ce, and Nd)}
\label{sec:neutroncapture}





{\bf Yttrium.} 
The derived [Y/Fe] abundances 
determined for the solar-neighbourhood sample, show sub-solar values, and scatter for values at a given metallicity for the thin-disk. The thick-disk stars have [Y/Fe] values that follow the lower envelope of the thin-disk trend. 
The six inner-bulge stars have scattered [Y/Fe] values consistent with the solar neighborhood [Y/Fe] trend and range. 

{\bf Cerium.}  
Just like in the case of the trend for yttrium, the thick-disk stars follow lower envelope of the thin-disk trend for the solar-neighbourhood stars \citep[see, e.g.,][]{rebecca:phd:23}. 
The trend from all the six inner-bulge stars are consistent with the solar-neighborhood, thick-disk [Ce/Fe] trend for low metallicities, and follow the general solar-neighborhood trend for the metal-rich stars.

{\bf Neodymium.}  
The [Nd/Fe] determinations for the thick-disk stars in the solar-neighbourhood sample again follow the lower envelope of the thin-disk. 
The inner-bulge stars follow the solar-neighborhood trend.

\subsubsection{Element with $50/50$ s- and r-process contributions (Yb)}

{\bf Ytterbium.}  
There is a large scatter ($\sim$ 0.5 dex) in the [Yb/Fe] values for the solar-neighborhood populations. 
Although there is a large scatter, the general picture that the thick-disk and the thin-disk [Yb/Fe] values more or less overlap is very similar to what \citet{rebecca:phd:23} find for praseodymium, Pr.  This element too has a 50/50 contribution of s-  and p-process origin in the solar isotopic mixture \citep{bisterzo:14}. 
We could determine the [Yb/Fe] abundances for only four metal-rich stars in the inner-bulge, and they are consistent with the solar-neighborhood values in the super-solar metallicity range.

\section{Discussion}
\label{sec:discussion}

Chemodynamic characterisation of the stellar populations in the inner-bulge is essential for disentangling their formation and evolutionary scenarios. They have, however, escaped a detailed chemical study due to the dust extinction along the line-of-sight. A differential chemical study comparing the inner-bulge populations with those in the solar neighbourhood, would reveal similarities and differences in environmental influences on, for example, the star-formation histories and processes, gas-flow patterns, evolutionary time scales, and stellar evolutionary processes in the different populations \citep[see, e.g.,][]{friske:23}.  

\subsection{The Bulge}
\label{sec:bulge}

The large structure referred to as the Milky Way bulge \citep[within $10^\circ$ of the Galactic Center;][]{barbuy:18} has successfully been chemically investigated in recent years. At least two components in the metallicity distribution function are found in previous studies \citep[e.g.][]{Hill:2011,Rojas-Arriagada:2017,Rojas-Arriagada:2020,Johnson:2022}, with a consensus on the metal-rich peak at $\sim +0.3$ dex to be associated with the barred bulge.
The physical origins and the number of the metal-poor peaks is, however, debated \citep{bland:16,Rojas-Arriagada:2020,Johnson:2022}. 

There is a consensus that the chemistry of the stars in the bulge is similar to that of the local thick-disk when it comes to the $\alpha$ elements,  such as Mg, Si, and Ca \citep{melendez:08,ryde:10,johnson:14,jonsson:17,bensby:17,lomaeva:19,Nieuwmunster:2023}, or at least following the upper envelope of the thick-disk trends. These elements are good probes for modelling the star-formation history of a stellar population \citep[see, e.g.,][]{grisoni:17,Matteucci:2021}. 

For the emerging picture with the bulge being the bar seen edge-on, these trends might be expected if the general picture of the $\alpha$ trends are extrapolated inward from the findings from the APOGEE survey \citep{hayden:15,SH:20}. These trends show a thick-disk dominating with a small scale-length and a metal-rich disk with low alphas that is getting progressively more metal-rich the smaller the galactocentric radius, reflecting the metallicity gradient. 

\subsection{The Inner-Bulge}
\label{sec:inner}

Chemical studies of the stellar populations in the inner-bulge regions ($|b|< 2^{\circ}$, or $\sim\pm 300$\,pc in projected distance from the Galactic Center) are, however, sparse. A few chemical studies have revealed
an $\alpha$-element trend versus metallicity again resembling that of the local thick-disk \citep[e.g.][]{Schultheis:2020,Nieuwmunster:2023}. The Mg, Si, and Ca trends in \citet{Nieuwmunster:2023} follow the thick-disk trends but sample more of the most metal-rich stars, a fact that was already seen in \citet{ryde:15} with a metal-rich population with thick-disk alpha trends in the Galactic Center populations. The stars investigated in \citet{Schultheis:2020}, situated within 150 pc of the Galactic Center, show a similar Mg, Si, Ca, and O trends to the Galactic bulge.

The very inner regions, the region within 200 pc (($|b|< 1.5^{\circ}$), is dominated by the nuclear stellar disk (NSD), whose chemical history might be expected to be different depending on formation scenario. We can assume that the NSD could have had intense star-formation dominated by outwards moving star-forming regions  \citep{friske:23}. These regions have been in the form of nuclear rings which have been fed by strong gas inflows driven by  the surrounding bar and funnelled from the bar tips  \citep{friske:23}. Different inflow scenarios distribute heavy elements very differently in the NSD with a possible pile-up toward the inner radial parts of the NSD,  \citep[for predictions, see][]{friske:23}.  

The NSD is a flattened structure with a stellar density-profile falling-off rapidly \citep[see][]{nishiymana:13,wegg:13,bland:16}. Our stars presented here all lie in a field  along the northern minor axis at ($l = 0^\circ$, $b = +1^\circ$), which is too far away from the plane to be considered belonging to the NSD. Our six stars can, therefore, be considered to be in the inner-bulge but not in any of the inner structures found there, especially the NSD. Stars closer to the center, in the NSD and nuclear star cluster (NSC), will be presented in a forthcoming paper. 

A dimension that is lacking in earlier discussions are elemental trends from other species than the $\alpha$ elements. All elements are formed from a unique mixture of different nucleosynthetic networks, acting at vastly different time scales. Therefore, by investigating a range of elements belonging to different classes of elements, new insights can be gained. Especially the
neutron-capture elements are distinct from the lighter elements. \citet{manea:23} illustrate clearly the importance of considering the neutron-capture elements when chemically characterizing stars. Among the 21 elements we have investigated here for the inner-bulge stars, we have representatives of the $\alpha-$, odd-Z, iron-peak, transition elements between iron-peak and
neutron-capture elements,  neutron-capture elements (i.e. weak-s, s-process dominated elements, r-process dominated elements), and fluorine, with its unique formation history \citep{Ryde:2020}.






Overall, we find a clear similarity and consistency between the element trends for all 21 elements, beside that of fluorine, for the stellar populations in  the inner-bulge, on one hand, and the local thick-disk and metal-rich thin-disk (inner disk sequence), on the other. The enhancements in the $\alpha$-element, as well as in Al, K, Sc, Ti, Co, and V abundances are strikingly similar for the metal-poor stars, following the local thick-disk trends. These elements mostly also follow the upper envelope of the local thin-disk at high metallicities. These similarities and consistencies are also found in the s-process dominated elements Y and Ce, where the thick-disk trends follow the lower envelope of the thin-disk scatter or 'cloud' at $-0.5<$\feh$<0$, see \cite{Nandakumar:2024}. Also, the weak-s process element, Cu, shows a great similarity between the stars in the local thick-disk and inner-bulge stars. Nd also has a dominant s-process origin, but to a lesser degree than for Ce. The trends for the stars in the inner-bulge and the local thick-disk are consistent with coming from a similar population for Nd also. These facts all strengthens the similarity between the inner-bulge and the local thick-disk in a new dimension, namely the neutron-capture dimension \citep{manea:23}. There is also a large similarity between the trends for the iron-peak elements (Cr, Mn, and Ni) of the inner-bulge stars and the solar neighbourhood stars. 
In summary, the hypothesis that the two populations are similar chemically can, therefore, not be rejected. 

\begin{figure}
  \includegraphics[width=\columnwidth]{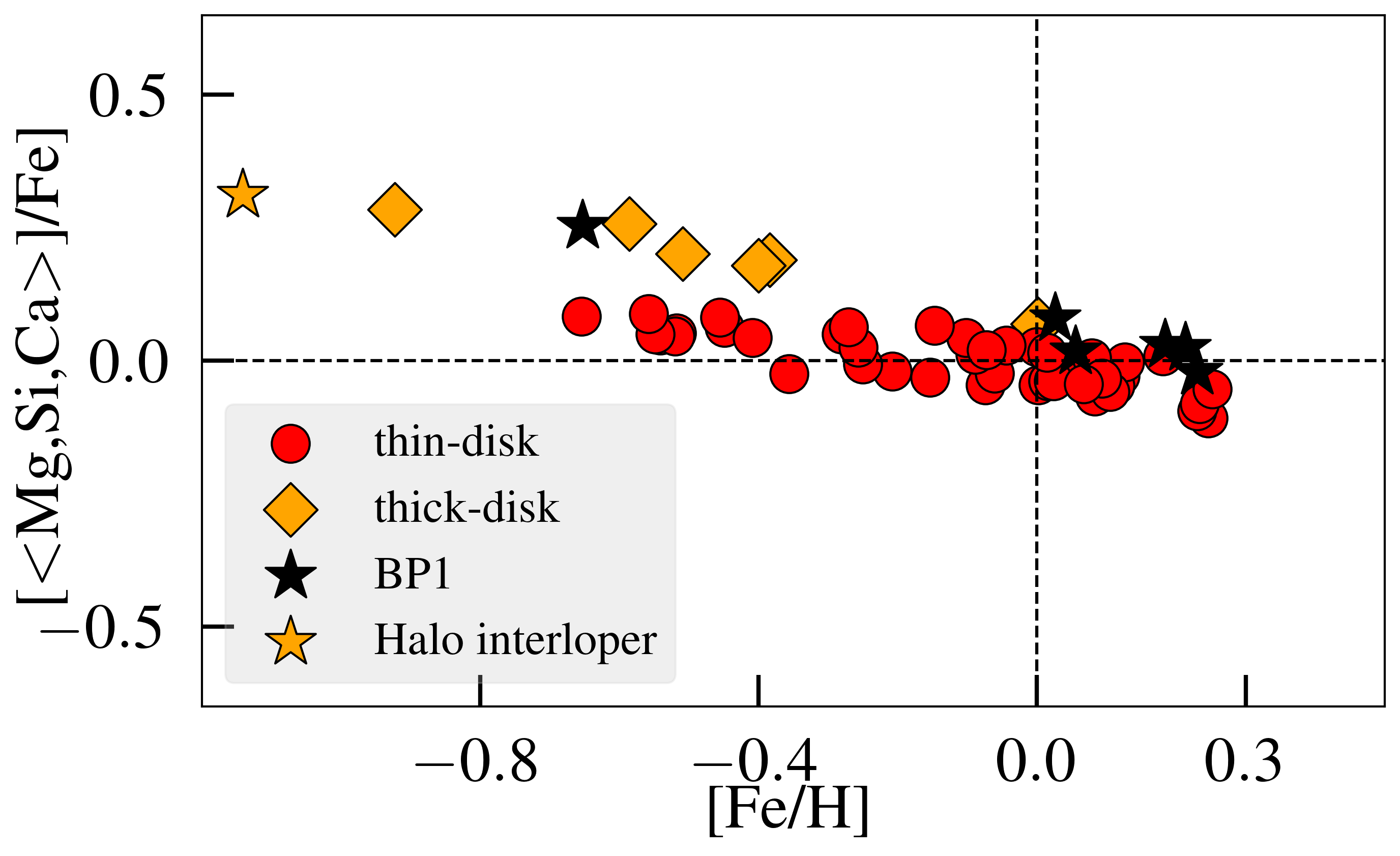}
  \caption{Mean abundances of the three $\alpha$ elements, Mg, Si, and Ca, versus [Fe/H] for 50 M giants in the solar neighborhood reference sample (red circles and orange diamonds) and the seven inner-bulge stars (black stars).  }
  \label{fig:alpha_mean}%
\end{figure}

In Figure \ref{fig:alpha_mean} we show the similarity between the local thick-disk plotted with orange diamonds for a mean of the three well determined $\alpha$ elements (Mg, Si, and Ca), and the inner-bulge stars plotted as black stars. The metal-rich inner-bulge stars follow the upper envelope of the local metal-rich, thin-disk. The precision of the data is also clear in the plot. Using the cosmological zoom simulation {\small VINTERGATAN} \citep[][]{Agertz2021}, \citet{Renaud2021} demonstrated that an upper envelope in $[\alpha/{\rm Fe}]$ trends can be associated with star formation in dense gas clouds where star formation is `fast' i.e. short gas depletion times \citep[see also][]{Clarke2019}. In these dense environments, stellar feedback is less efficient at disrupting the natal clouds, leading to $\alpha$-rich ejecta from core-collapse supernovae rapidly being reincorporated into the star formation process, hence enhancing $[\alpha/{\rm Fe}]$ in newly formed stellar populations at a relatively constant $[{\rm Fe/H}]$. It remains to be understood whether the enhancement in our metal-rich sample of stars \citep[see also][]{thorsbro:2020} reflects the past star formation conditions of the inner-bulge, or whether this population of stars formed at greater galactocentric radii in dense gas, massive clouds that migrated to the center due to dynamical friction \citep[e.g.][]{Immeli2004,Elmegreen2008,Agertz2009}.

The only clear differences between the populations are {\it (i)} the distribution of metallicities of the stellar populations, being more metal-rich in the inner-bulge regions and {\it (ii)} the fluorine trend with metallicity. We refrain from drawing any conclusions from the higher fluorine trend at high metallicities, due to the temperature sensitive nature of the lines. A large set of stars in the inner-bulge region for a range of metallicities would be needed to draw more conclusions.  However, an explanation that would increase the fluorine trend at higher metallicities, and especially at supersolar metallicities, is a contribution to the cosmic budget for fluorine from novae, as suggested by \citet{Spitoni:2018}. This process increases its yields with metallicity at supersolar metallicities \citep[see models in][]{Ryde:2020}. Whether or not  novae could contribute at supersolar metallicities, is currently not known since the novea yields are highly uncertain \citep{Spitoni:2018}.

\begin{figure*}
  \includegraphics[width=\textwidth]{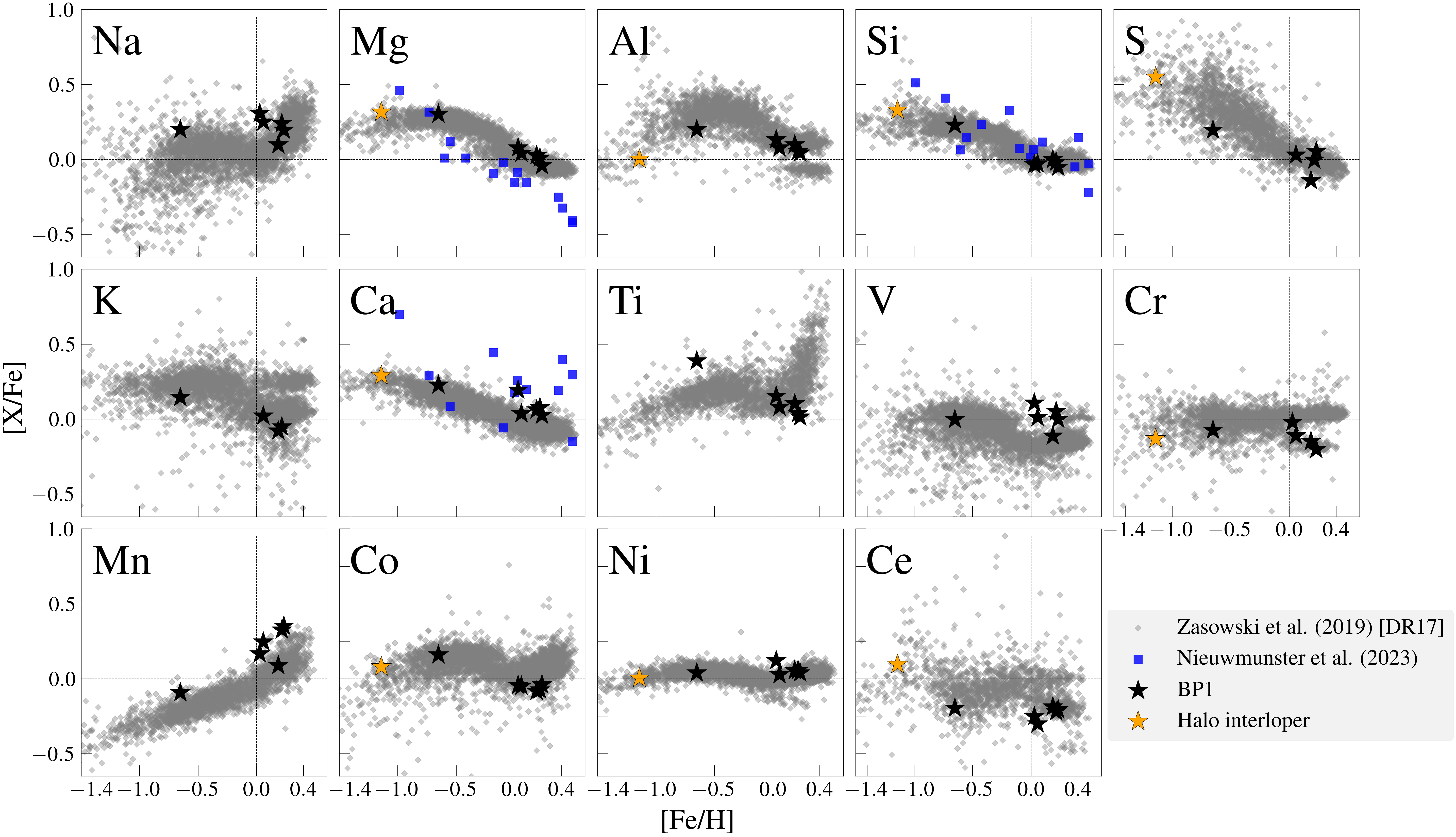}
  \caption{Elemental abundances of 14 elements versus metallicity for inner Milky Way stars (grey squares) investigated in \cite{Zasowski:2019} selected from APOGEE DR17 catalog, and inner-bulge stars in this work (black star markers). Abundances of Mg, Si and Ca for stars located in the BP1 field in \cite{Nieuwmunster:2023} are shown as blue squares.  }
  \label{fig:all_apogee_trend}%
\end{figure*}



We also determined abundances of all 21 elements with the new sets of stellar parameters determined with the thin-disk oxygen assumption as described in the section~\ref{sec:thin_parameters}. This provides us with an estimate of the abundance variations due to the change in stellar parameters from the oxygen assumption. For the metal-rich stars ([Fe/H] $>$ 0), Mg, Al, Si, Ca, Zn, and Yb exhibit the smallest difference ($<$ 0.06 dex) and the rest of the elements have differences less than 0.18 dex. The most metal-poor star, BP1$\_$5, with the largest difference in \teff\, (150 K) and \logg\, (0.25 dex), exhibit difference of $\sim$0.27 dex for cerium, and much lower differences ($<$ 0.18 dex) for all other elements. The largest differences in the case of metal-rich stars are found for fluorine (0.18 dex) and scandium (0.14 dex). This can, indeed, be attributed to the \teff-sensitivity of the HF and Sc{\sc i} lines, from which we determined the abundances (see also sections ~\ref{sec:fluorine} and ~\ref{sec:ironpeak}). Such large differences would make the fluorine abundances consistent with the local thin-disk abundances, hence the reason for refraining from drawing any conclusions from the higher fluorine trend at high metallicities.



\subsection{Comparison with inner Milky Way stars}
\label{sec:IMWcompare}

Next, we compare our abundance trends with those from the inner Milky Way stellar sample of \cite{Zasowski:2019} but using the abundances provided in the APOGEE DR17 catalog. We also include the Mg, Si and Ca abundance trends determined from CRIRES K-band spectra for 16 K and M giants in the $(l,b)=(0,+1^{\circ})$ field in \cite{Nieuwmunster:2023}. 

\cite{Zasowski:2019} selected $\sim$ 4000 stars with 3600 K $\leq$ Teff $\leq$ 4500 K and -0.75 $\leq$ log g $\leq$ 3.5 in APOGEE DR14/DR15 that are at galactocentric distances, R $\leq$ 4.0 kpc,  and investigated the elemental abundance trends of 11 elements. We thus selected the same set of stars from the APOGEE DR17 catalog and compared our elemental abundance trends for 14 elements (3 more available in APOGEE DR17) in Figure~\ref{fig:all_apogee_trend}, where the Mg, Si and Ca abundance trends from \cite{Nieuwmunster:2023}, mentioned above, are also plotted. 

\cite{Zasowski:2019} found the lack of a bimodality in the [$\alpha$/Fe] distributions at subsolar metallicity suggesting that the abundances are dominated by a single chemical evolutionary sequence. They also found inner Milky Way abundances for several non-$\alpha$-elements to be similar to the solar radius stars in APOGEE that belong to the high-$\alpha$ ([$\alpha$/M] $>$ 0.12) sample. Our findings agree with this, and more importantly we add more chemical species from other nucleosynthetic channels, not investigated before in the inner-bulge region. We show that the  trends of these newly added elements also exhibit values similar to the solar neighborhood thick-disk stars. 

In Figure~\ref{fig:all_apogee_trend}, we find excellent agreements in the case of elements with the least scatter in APOGEE: Mg, Si, Ca, and Ni. Both Na and S in APOGEE show good agreement with our abundance values though APOGEE show large scatter at sub-solar metallicities. Significant differences are seen for Ti, V, Cr, and Mn. Our Ti abundances are significantly better and agree with the predictions from theoretical chemical evolution models compared to the N-shaped trend in APOGEE. We also note that \cite{Hawkins:2016} have improved the abundance ratios of [Ti/Fe] and [V/Fe] (with respect to those in APOGEE DR12) with the reanalysis of APOGEE spectra using the BACCHUS pipeline through a careful line selection. For Mn, we have used NLTE corrections based on calculations by \cite{amarsi:20} whereas NLTE corrections for Mn are not included in APOGEE. We have also included hyperfine splitting information for the Mn lines from \cite{montelius:21} while it is not clear whether APOGEE linelist includes this information for Mn. For Al, K, and Cr, there is a bifurcation at high metallicities for APOGEE, and \cite{Zasowski:2019} have shown that Al, K, Cr, Mn, and Co in APOGEE DR14/DR15 exhibit \teff-dependencies for stars with \teff $<$ 3800 K. We note that we do not see \teff\, dependence for these elements based on the comparison of our solar neighborhood reference sample trends with the elemental trends from the GILD survey (J\"onsson et al. in prep) which is an optical survey of $\sim$500 warmer K-giants in the solar neighbourhood \citep{Nandakumar:2024}. In the case of Ce, our inner-bulge abundances lie in the lower envelope of the APOGEE Ce trend, but they are consistent with the solar neighbourhood reference sample as shown in Figure~\ref{fig:all_gild_trend}. This difference in Ce abundances could be due to the choice of different Ce\,{\sc ii} lines in APOGEE and this work, or due to the systematic differences in stellar parameters (see the paragraph below).  

In general, most of the disagreements/differences with APOGEE may be attributed to a combination of different factors like differences in spectral resolution, careful line selection, line lists, pipelines etc. As mentioned before, this has been shown in \cite{Hawkins:2016} and \cite{Hayes:2022} where they reanalyse APOGEE spectra using the independent pipeline BACCHUS along with an improved linelist and careful line selection. This has improved the abundance trends of elements like Na, Ti, S, V, and Ce and has led to measuring elements that APOGEE fails, such as P, Cu, and Nd. In this work, we determine stellar parameters self consistently by spectral synthesis from IGRINS spectra using SME. A comparison between our stellar parameters and those from APOGEE for 44 solar neighbourhood stars, which are also in APOGEE, in \cite{Nandakumar:2023} have shown that there are systematic differences of $\sim$ 100 K in \teff, 0.3 dex in \logg, and 0.2 kms$^{-1}$ in $\xi_\mathrm{micro}$. This could in turn lead to systematic differences in elemental abundances. Furthermore, for elements such as Mg, Ca, S, Ti, and Ni, we were able to choose more lines from K-band owing to the broader wavelength coverage of IGRINS spectra (full H- and K-bands). This could also lead to differences in the abundance trends between this work and APOGEE. In addition, IGRINS has nearly two times higher resolution than APOGEE. This is crucial in capturing blends and uncertain continuum in the crowded spectra of cool metal-rich stars, and thus lead to more accurate abundances from IGRINS spectra.



\cite{Nieuwmunster:2023} determined the three $\alpha$-element abundances from the few K-band lines available in CRIRES spectra. Though both our work and \cite{Nieuwmunster:2023} use the same linelist, only two Mg, and Ca lines, and three Si lines are in common between our work and that of \cite{Nieuwmunster:2023}. While there is significant scatter in the Si abundances for stars in \cite{Nieuwmunster:2023}, there is generally a good agreement with the abundances of our stars. The Mg abundances have the least scatter but are lower compared to our abundances. The Ca abundances in \cite{Nieuwmunster:2023} are elevated for the supersolar metallicities and there is a large scatter as well. When we determine the Mg and Ca abundances from the same two lines used in \cite{Nieuwmunster:2023}, the Mg abundances decrease \citep[see also Figure 9 in ][]{Nandakumar:2023} and the Ca abundances increase for the supersolar metallicity stars in our sample, consistent with \cite{Nieuwmunster:2023}. Hence, in addition to self-consistently determining stellar parameters from the same spectra, the ability to choose better H- and K-band lines have improved the abundances in this work. This emphasize the importance of an instrument like IGRINS, that covers the full H- and K-band wavelength regimes, in spectroscopic studies of stellar populations in the Milky Way. 



\section{Conclusions}
\label{sec:conclusion}

The chemical evolution of the Galactic bulge as a function of Galactic latitude is of great interest for the study of the Milky Way, and the problem is likely to be one of increasing complexity as sample sizes, and number of elements, increase.   Especially, the inner, dust-obscured regions, where most of the mass lies, holds important and likely interesting information. The innermost degree also contains the Central Molecular Zone, Nuclear Stellar Disk, and Nuclear Star Cluster, as well as the stellar-density peak of the inner stellar halo.  In some of these locations, star formation may be ongoing, and stars formed within the last few Gyr may contribute to the population \citep[see e.g.][]{Thorsbro:2023}.
These regions have, however, not been chemically explored much due to heavy optical extinction along the line-of-sight. To disentangle these different structures, a chemical characterisation of the stellar populations there is needed.  Infrared spectroscopic data at the highest quality and spectral resolution is required to achieve high precision and accuracy. The Immersion GRating INfrared Spectrograph (IGRINS) meets these requirements, observing the full H and K bands.

Taking advantage of the extraordinary high spectral resolution and wavelength range of IGRINS, we present a spectral analysis and the first heavy element trends, and Flourine measurements, for 7 M-giants lying at ($l,b$)=(0,+1$^{\circ}$).  Six stars in our sample are inner-bulge stars, probably in the vicinity of, but not in, the Nuclear Stellar Disk. We present a differential investigation of 21 elemental trends versus metallicity at high precision. We compare with derived abundances for stars in the solar neighborhood region of the same spectral types (here, M giants). In order to be strictly differential and minimize any systematic uncertainties, we use the same instrument with the same wavelength range, spectral resolution, pipelines, and analysis techniques  in the determination of stellar parameters and elemental abundances. The large number of elements offers new dimensions in chemical space, beyond the metallicity and the commonly studied $\alpha$ elements. This provides new insights \citep[see, e.g,][]{manea:23} and potentially can break degeneracies. There will be a tension among results as detailed analyses based on spectrum synthesis of individual spectra bump up against the results of very large programs.   


We find  that the stars in the inner bulge, $1^\circ$ North of the Galactic Center, are strikingly homogeneous with the local thick-disk (i.e. the solar neighbourhood ‘inner disk sequence’ as referred by \citealt{Haywood:2013}) in the chemical trends for 21 of the elements that includes all nucleosynthetic channels, i.e.
$\alpha$, odd-Z, iron-peak, transition elements between iron-peak and
neutron-capture elements, and neutron-capture elements (i.e. weak-s, s-process dominated elements, r-process dominated elements). The hypothesis that the two populations are similar chemically cannot be rejected. 

The only clear differences between the populations are, first, that  we note that we are efficiently detecting metal-rich stars in the inner-bulge regions and fewer stars at medium metallicities and, second, that  the fluorine trend with metallicity {\it could be} different from the trend in the solar neighbourhood. Although, premature, we can note that a way to increase the fluorine abundance in a stellar population is a contributions to the cosmic budget for fluorine from novae, as suggested by \citet{Spitoni:2018}. If this would indeed be the case is very unclear since the nova yields are highly uncertain \citep{Spitoni:2018}. In any case, the fluorine trend needs more attention and a more robust trend needs to be established for the stellar populations in the inner-bulge structures. We also note that, among the few stars we observe in the inner-bulge, we have observed 1/6 metal-poor stars (\feh$<-0.5$). 

We have demonstrated that a precise abundance analysis of as many as 21 elements of stars in the inner-bulge is possible with the IGRINS spectrometer. The remarkable capabilities of this instrument -- access to both the H and K infrared bands with complete spectral coverage at resolution $R=45,000$ will not be matched by the next generation of large scale infrared high resolution spectroscopy surveys, which utilize only the H band at resolution $R=18,000-23,000$.  For these next generation surveys, study of a number of interesting heavy elements, and Flourine, will be impossible.  Our investigation emphasizes the need for continued work on these very interesting inner galaxy populations using the currently most powerful infrared spectrograph for high resolution spectroscopy, IGRINS. 

Our data presented here were observed with 
$2.7$~meter (107-inch) Harlan J. Smith Telescope at McDonald Observatory in May-July 2016. The next step is to observe the structures in the Galactic Center region, which is possible with larger telescopes. A forthcoming paper will present 21 elemental trends for stars in the Nuclear Star Cluster.


\begin{acknowledgements}
We thank the anonymous referee for the constructive comments and suggestions that improved the quality of the paper. G.N.\ acknowledges the support from the Wenner-Gren Fo undations (UPD2020-0191 and UPD2022-0059) and the Royal Physiographic Society in Lund through the Stiftelsen Walter Gyllenbergs fond. N.R.\ acknowledge support from the Royal Physiographic Society in Lund through the Stiftelsen Walter Gyllenbergs fond and Märta och Erik Holmbergs donation. OA acknowledges support from the Knut and Alice Wallenberg Foundation and the Swedish Research Council (grant 2019-04659). B.T.\ acknowledges the financial support from the Wenner-Gren Foundation (WGF2022-0041). E.A.\ acknowledge support from US National Science Foundation under grants AST18-15461 and AST23-07950. This work used The Immersion Grating Infrared Spectrometer (IGRINS) was developed under a collaboration between the University of Texas at Austin and the Korea Astronomy and Space Science Institute (KASI) with the financial support of the US National Science Foundation under grants AST-1229522, AST-1702267 and AST-1908892, McDonald Observatory of the University of Texas at Austin, the Korean GMT Project of KASI, the Mt. Cuba Astronomical Foundation and Gemini Observatory.
This work is based on observations obtained at the international Gemini Observatory, a program of NSF’s NOIRLab, which is managed by the Association of Universities for Research in Astronomy (AURA) under a cooperative agreement with the National Science Foundation on behalf of the Gemini Observatory partnership: the National Science Foundation (United States), National Research Council (Canada), Agencia Nacional de Investigaci\'{o}n y Desarrollo (Chile), Ministerio de Ciencia, Tecnolog\'{i}a e Innovaci\'{o}n (Argentina), Minist\'{e}rio da Ci\^{e}ncia, Tecnologia, Inova\c{c}\~{o}es e Comunica\c{c}\~{o}es (Brazil), and Korea Astronomy and Space Science Institute (Republic of Korea).
\end{acknowledgements}

\vspace{5mm}
\facilities{107-inch Harlan J. Smith Telescope at McDonald Observatory (Immersion GRating INfrared Spectrograph \citep[IGRINS;][]{Yuk:2010,Wang:2010,Gully:2012,Moon:2012,Park:2014,Jeong:2014}).}

\software{SME \citep{sme,sme_code}, IRAF \citep{IRAF}, TOPCAT (version 4.6; \citealt{topcat}); Python (version 3.8) and its packages ASTROPY (version 5.0; \citealt{astropy}), SCIPY \citep{scipy}, MATPLOTLIB \citep{matplotlib} and NUMPY \citep{numpy}.}

\bibliography{references}{}
\bibliographystyle{aasjournal}






\end{document}